\documentclass[a4paper,12pt]{article}
\usepackage{a4}
\usepackage{amsbsy}
\usepackage{amssymb}
\usepackage{amsmath}
\usepackage{epsfig}
\usepackage{fullpage}

\begin{document}

\newtheorem{theorem}{Theorem}
\newtheorem{lemma}{Lemma}

%%Useful symbols%%%%%%%%%%%%%%%%%%%%%%%%%%%%%%%%%%
\def\CA{{\cal A}}
\def\CB{{\cal B}}
\def\CC{{\cal C}}
\def\CD{{\cal D}}
\def\CE{{\cal E}}
\def\CF{{\cal F}}
\def\CG{{\cal G}}
\def\CH{{\cal H}}
\def\CI{{\cal I}}
\def\CJ{{\cal J}}
\def\CK{{\cal K}}
\def\CL{{\cal L}}
\def\CM{{\cal M}}
\def\CN{{\cal N}}
\def\CO{{\cal O}}
\def\CP{{\cal P}}
\def\CQ{{\cal Q}}
\def\CR{{\cal R}}
\def\CS{{\cal S}}
\def\CT{{\cal T}}
\def\CU{{\cal U}}
\def\CV{{\cal V}}
\def\CW{{\cal W}}
\def\CX{{\cal X}}
\def\CY{{\cal Y}}
\def\CZ{{\cal Z}}

%macros
\newcommand{\todo}[1]{{\em \small {#1}}\marginpar{$\Longleftarrow$}}
\newcommand{\labell}[1]{\label{#1}\qquad_{#1}} %{\label{#1}} %
\newcommand{\bbibitem}[1]{\bibitem{#1}\marginpar{#1}}
\newcommand{\llabel}[1]{\label{#1}\marginpar{#1}}
\newcommand{\restrict}[1]{\lfloor_{#1}}

% macros for the conical defect paper
\newcommand{\sphere}[0]{{\rm S}^3}
\newcommand{\su}[0]{{\rm SU(2)}}
\newcommand{\so}[0]{{\rm SO(4)}}
\newcommand{\bK}[0]{{\bf K}}
\newcommand{\bL}[0]{{\bf L}}
\newcommand{\bR}[0]{{\bf R}}
\newcommand{\tK}[0]{\tilde{K}}
\newcommand{\tL}[0]{\bar{L}}
\newcommand{\tR}[0]{\tilde{R}}

\newcommand{\btzm}[0]{BTZ$_{\rm M}$}
\newcommand{\ads}[1]{{\rm AdS}_{#1}}
\newcommand{\ds}[1]{{\rm dS}_{#1}}
\newcommand{\eds}[1]{{\rm EdS}_{#1}}
\newcommand{\sph}[1]{{\rm S}^{#1}}
\newcommand{\gn}[0]{G_N}
\newcommand{\SL}[0]{{\rm SL}(2,R)}
\newcommand{\cosm}[0]{R}
\newcommand{\hdim}[0]{\bar{h}}
\newcommand{\bw}[0]{\bar{w}}
\newcommand{\bz}[0]{\bar{z}}
\newcommand{\be}{\begin{equation}}
\newcommand{\ee}{\end{equation}}
\newcommand{\bea}{\begin{eqnarray}}
\newcommand{\eea}{\end{eqnarray}}
\newcommand{\pat}{\partial}
\newcommand{\lp}{\lambda_+}
\newcommand{\bx}{ {\bf x}}
\newcommand{\bk}{{\bf k}}
\newcommand{\bb}{{\bf b}}
\newcommand{\BB}{{\bf B}}
\newcommand{\tp}{\tilde{\phi}}
\newcommand{\twoa}[4]{\left(\begin{array}{cc} #1 & #2 \\ #3 & #4
\end{array}\right)}
\newcommand{\threea}[9]{\left(\begin{array}{ccc} #1 & #2 & #3 \\ #4 &
#5 & #6 \\ #7 & #8 & #9 \end{array} \right)}
\hyphenation{Min-kow-ski}

%%Commonly used constants and symbols%%%%%%%%%%%%%%%%%%%%%%%%%
\def\apr{\alpha'}
\def\str{{str}}
\def\lstr{\ell_\str}
\def\gstr{g_\str}
\def\Mstr{M_\str}
\def\lpl{\ell_{pl}}
\def\Mpl{M_{pl}}
\def\varep{\varepsilon}
\def\del{\nabla}
\def\grad{\nabla}
\def\perp{\bot}
\def\half{\frac{1}{2}}
\def\p{\partial}
\def\perp{\bot}
\def\eps{\epsilon}

\newcommand{\tr}{\mathrm{tr}}

\def\NPB{{\it Nucl. Phys. }{\bf B}}
\def\PL{{\it Phys. Lett. }}
\def\PRL{{\it Phys. Rev. Lett. }}
\def\PRD{{\it Phys. Rev. }{\bf D}}
\def\CQG{{\it Class. Quantum Grav. }}
\def\JMP{{\it J. Math. Phys. }}
\def\SJNP{{\it Sov. J. Nucl. Phys. }}
\def\SPJ{{\it Sov. Phys. J. }}
\def\JETPL{{\it JETP Lett. }}
\def\TMP{{\it Theor. Math. Phys. }}
\def\IJMPA{{\it Int. J. Mod. Phys. }{\bf A}}
\def\MPL{{\it Mod. Phys. Lett. }}
\def\CMP{{\it Commun. Math. Phys. }}
\def\AP{{\it Ann. Phys. }}
\def\PR{{\it Phys. Rep. }}

%%%%%%%%%%%%%%%%%%%%%%%%%%%%%%%%%%%
% Erich's macros:
\renewcommand{\thepage}{\arabic{page}}
\setcounter{page}{1}

%\title
\rightline{hep-th/0305189}
\rightline{ITFA-2003-25}
\vskip 0.75 cm
\renewcommand{\thefootnote}{\fnsymbol{footnote}}
\begin{center}
\Large \bf  Factorization of Seiberg-Witten Curves and\\
Compactification to Three Dimensions
\end{center}
\vskip 0.75 cm

\centerline{{\bf Rutger Boels,\footnote{rhboels@science.uva.nl}
Jan de Boer,\footnote{jdeboer@science.uva.nl}
Robert Duivenvoorden\footnote{rjduiven@science.uva.nl}
and
Jeroen Wijnhout\footnote{wijnhout@science.uva.nl}
}}
\vskip .5cm
\centerline{\it Instituut voor Theoretische Fysica,}
\centerline{\it Valckenierstraat 65, 1018XE Amsterdam, The Netherlands}
\vskip .5cm

\setcounter{footnote}{0}
\renewcommand{\thefootnote}{\arabic{footnote}}

\begin{abstract}
\noindent We continue our study of nonperturbative superpotentials
of four-dimensional \mbox{${\cal N}=2$} supersymmetric gauge
theories with gauge group $U(N)$ on \mbox{$\mathbb{R}^3\times
S^1$}, broken to ${\cal N}=1$ due to a classical superpotential.
In a previous paper \cite{prevpaper} we discussed how the
low-energy quantum superpotential can be obtained by substituting
the Lax matrix of the underlying integrable system directly into
the classical superpotential. In this paper we prove algebraically
that this recipe yields the correct factorization of the
Seiberg-Witten curves, which is an important check of the
conjecture. We will also give an independent proof using the
algebraic-geometrical interpretation of the underlying integrable
system.
\end{abstract}
\newpage

\section{Introduction}
Despite its phenomenal success in predicting phenomena at high
energies, the theory of strong interactions still lacks
understanding at low energies. Recently progress has been made in
understanding the vacuum structure of supersymmetric $N=1$
extensions of $4$ dimensional gauge theories \cite{dv1, dv2},
which could provide a useful laboratory for studying $QCD$. As a
part of this ongoing effort we will study supersymmetric gauge
theories on $\mathbb{R}^3 \times S^1$. As argued in
\cite{seibergwitten3d}, various holomorphic data including the
vacuum structure and the value of the superpotential in each
vacuum are independent of the radius $R$ of the $S^1$. In
particular this means that we can recover the four-dimensional ($R
\to \infty$) vacuum structure from the three-dimensional (finite
$R$) field theory.

The quantum structure of the three-dimensional field theory (
$\mathcal{N} = 4$ supersymmetric gauge theory deformed to
$\mathcal{N}=2$ by a classical superpotential) turns out to be
easier to describe than the four-dimensional gauge theory. There
are two main reasons why this is the case. First of all there are
no additional light degrees of freedom that can condense. Secondly
there are no fractional instantons in the theories we consider.
This ultimately implies that we can find the quantum
superpotential by expressing the classical superpotential in the
right variables.

These right variables are provided by the integrable system that
underlies the Seiberg-Witten solution of ${\cal N}=2$
supersymmetric gauge theories in four dimensions \cite{gorsky} \cite{martinecwarner} \cite{nakatsu}. This integrable system takes on
a more direct physical meaning once we compactify the
four-dimensional theory down to three dimensions, as the moduli
space of ${\cal N}=4$ gauge theories in three dimensions coincides
with the complexified phase space of the integrable system. We
therefore proposed in \cite{prevpaper}, following earlier work
\cite{dorey1} \cite{dorey2}\cite{dorey3} , that one can obtain the quantum
superpotential simply by replacing the adjoint scalar $\Phi$ in
the classical superpotential by the Lax matrix $M$ of the
corresponding integrable system,
\begin{equation}
    \label{eq:conjecture}
    W_\mathrm{quantum} (\Phi) = W_\mathrm{classical} (M).
\end{equation}

We were able to show, in several cases, that this proposal
correctly reproduced the vacuum structure of the four-dimensional
supersymmetric gauge theories. That is, we were able to show that
at extrema of the superpotential, the Seiberg-Witten curve
factorizes precisely in the right way as it should according to
the four-dimensional analysis in
\cite{seibergwittencurvefactorization}. In the case where the
gauge group is $U(N)$, this factorization implies that the
Seiberg-Witten curve degenerates to a curve of genus $k<N-1$, and
the general form of this factorization is
\begin{equation}
    \label{eq:factor1}
    y^2 = P^2_N(x) - 4 \Lambda^{2N} = H^2_{N-k}(x) T_{2k}(x)
\end{equation}
\begin{equation}
    \label{eq:factor2}
    G^2_{n-k}(x) T_{2k}(x) = W'(x)^2 + f_{n-1}(x),
\end{equation}
where $P_N(x)$ is the $z$-independent part of $\det ( x - M(z))$
($z$ is the spectral parameter that appears in the Lax matrix
$M(z)$) and $H,T,G,f$ are polynomials of degree $N-k,2k,n-k,n-1$
respectively. The degree of the superpotential $W$ is $n+1$.

In this paper we will prove that (\ref{eq:conjecture}) reproduces
the correct factorization of the Seiberg-Witten curve
(\ref{eq:factor1}),(\ref{eq:factor2}). The proof is purely
algebraic and relies heavily on the (almost) tri-diagonal shape of
the Lax matrix. In section \ref{sec:alggeomproof} we will give a
short alternative proof using methods from the theory of
integrable systems and Riemann surfaces. We also briefly discuss
the number of flat directions of (\ref{eq:conjecture}), the
difference between the gauge groups $SU(N)$ and $U(N)$, and
possible generalizations to other gauge groups.

While this paper was being typed, \cite{hollowood} appeared, in
which yet another proof of the correctness of
(\ref{eq:conjecture}) is given.

\section{The setup}
 
As explained in the previous paper \cite{prevpaper} (section 3) we are interested in extrema of the potential ${\rm Tr} W(M)$ obtained by inserting the Lax matrix $M$ of the periodic Toda chain into the superpotential of the deformed ${\cal N} =2$ $U(N)$ super Yang-Mills theory. In this section we will review the exact statements to be proven and along the way fix some notation. 

The periodic Toda chain Lax matrix is given as \cite{lax}

\bea \label{deflax}
M & = &
 \begin{pmatrix}
  p_1        &   \Lambda^2 e^{(q_1-q_2)/2}   &   0    & \dots & \Lambda^2 e^{(q_N-q_1)/2} z         \\
  \Lambda^2 e^{(q_1-q_2)/2}            & p_2  &  \Lambda^2 e^{(q_2-q_3)/2}   & \dots & 0         \\
   0            &     \Lambda^2 e^{(q_2-q_3)/2}    & p_3 & \dots & 0         \\
   .            &    .    &  .     & \dots & \Lambda^2 e^{(q_{N-1}-q_N)/2}   \\
   \Lambda^2 e^{(q_N-q_1)/2} z^{-1} &    .    &  .     &   \Lambda^2 e^{(q_{N-1}-q_N)/2}   & p_N
 \end{pmatrix} \nonumber \\
 & \equiv &
 \begin{pmatrix}
  \phi_1        &   y_1   &   0    & \dots & y_{0} z         \\
   y_1            & \phi_2  &  y_2   & \dots & 0         \\
   0            &    y_2    & \phi_3 & \dots & 0         \\
   .            &    .    &  .     & \dots & y_{N-1}   \\
   y_{0} z^{-1} &    .    &  .     &   y_{N-1}   & \phi_N
 \end{pmatrix}.
\eea

Note that we have changed the definition of the $y_i$ with respect to our previous article in order to declutter notation in this article. Also, we have changed to a symmetric Lax matrix representation, which is related to the one in \cite{prevpaper} by conjugation by a diagonal matrix. Another representation, which will be used in the algebraic proof in the next section, can be obtained by using the embedding $\widehat{gl}_N$ in $gl_{\infty}$,

\be \label{inflax}
\tilde{M}  = 
 \begin{pmatrix}
\ldots & \ldots & \ldots & \ldots &   &  \\
\ldots & \phi_N & y_0    &   0     &   0     &  \\
\ldots & y_0    & \phi_1 &  y_1    &   0	    & \ldots  \\
\ldots &  0     &  y_1	& \phi_2  &  y_2    & \ldots  \\
 &  0     &    0   &  y_2    & \phi_3  & \ldots  \\
&     &    \ldots   &   \ldots     & \ldots  & \ldots
 \end{pmatrix}.
\ee

The Lax matrix is then an infinite tridiagonal matrix $\tilde{M}$ with period $N$. If we 
define a shift operator $Z$ of order $N$ by the equation $Z=D^N$, where $D$ is the shift 
operator of order one with matrix elements $D_{ij}=\delta_{i,j-1}$, then clearly
$\tilde{M}$ commutes with $Z$. By considering the action of $\tilde{M}$ on vectors
that are eigenvectors of $Z$ with eigenvalue $z$, so that they have only $N$ independent
components, we recover (\ref{deflax}). 

Another ingredient in the setup is the operator ${\cal L}$ which will act on integer powers of the Lax matrix $M$; $M$ depends on a spectral parameter $z$ and can therefore be expanded as
\be M^n = \sum_r z^r M^n_{(r)} \ee
Furthermore, define $M^n_+$ as the sum of the upper diagonal part of $M^n_{(0)}$ plus
$\sum_{r<0} z^r M^n_{(r)}$, $M^n_-$ as the sum of the lower diagonal part of
$M^n_{(0)}$ plus $\sum_{r>0} z^r M^n_{(r)}$, and finally $M^n_0$ as the diagonal
part of $M^n_{(0)}$. In terms of these, $\CL$ acts as,
\be \label{defll}
{\cal L}(M^n) \equiv M^n_- + M^n_0 - M^n_+ .
\ee
Note that in the infinite matrix representation $\tilde{M}^n_-$, $\tilde{M}^n_0$ and
$\tilde{M}^n_+$ are simply the lower triangular, diagonal and upper triangular part
of $\tilde{M}$ respectively. The linear operator (\ref{defll}) has a natural interpretation
in terms of the affine root system of $\widehat{gl}_N$, it flips the signs of all positive
roots but does nothing to the negative roots. For every $1<n<N-1$ these ${\cal L}(M^n)$ generate the independent isospectral flows on the phase space of the system.
\be \label{commflow}
\frac{\partial M}{\partial t_n} = [M,{\cal L}(M^n)].
\ee
These flows commute and clearly preserve the quantities ${\rm Tr}(M^l)$. The latter are
the action variables of the periodic Toda chain, while the times $t_n$ are essentially
the angle variables, and altogether this demonstrates the integrability of the 
Toda chain. 

The spectral curve associated to the Lax matrix $M$ is defined by the equation 
\be \label{defspec}
\det (x - M) \equiv P_N(x) + (-1)^N(z+\Lambda^{2N} z^{-1}) = 0
\ee
and is identified with the Seiberg-Witten curve of the four-dimensional ${\cal N}=2$ 
theory. 

The flows (\ref{commflow}) preserve the product $\prod_{i=0}^{N-1} y_i^2 = \Lambda^{2N}$,
and in \cite{prevpaper} we enforced this condition using a Lagrange multiplier
field. Since the flows do not affect $\Lambda$, we will not write this Lagrange
multiplier term explicitly in the remainder, but freely replace the product
$\prod_{i=0}^{N-1} y_i^2$ by $\Lambda^{2N}$ whenever appropriate. 

The flows (\ref{commflow}) also preserve ${\rm Tr}(M)$. 
If we take $M$ to be traceless, we are discussing
the $SU(N)$ theory, while if we include the trace of $M$ we are discussing the $U(N)$
theory. In the latter case, the moduli space also includes the expectation values
of an extra complex scalar field which is obtained from the four dimensional
diagonal $U(1)\subset U(N)$ gauge field after compactifying on a circle and dualizing the
remaining three dimensional gauge field. This extra complex scalar is completely decoupled
from the discussion since it cannot appear in the superpotential and we will ignore it
in the remainder. In the remainder of the paper we will assume the gauge
group is $U(N)$ and briefly comment on the $SU(N)$ in section~5.

\subsection{Reduction of the integrable system}

We are interested in the extrema of the superpotential 
${\rm Tr} W(M) \equiv {\rm Tr} \sum_{i=1}^{n+1} \frac{g_i}{i} M^i$
of maximum power $n+1$. These are found by differentiating the superpotential
with respect to the coordinates and momenta of the Toda chain, and by putting these
equal to zero. Equivalently, the Poisson brackets of ${\rm Tr}W(M)$ with the coordinates
and momenta should vanish. Since ${\rm Tr}W(M)$ is a linear combination of action
variables ${\rm Tr}(M^k)$, vanishing of the Poisson brackets implies that the flow
generated by the action ${\rm Tr}W(M)$ should have a stationary point, which in turn
translates into the two matrix equations 
\be \label{weq1}
[M,W'(M)_+ - W'(M)_-]=0, \qquad W'(M)_0=0.
\ee
The second equation $W'(M)_0=0$ appears because ${\rm Tr}(M)$ does not generate
a flow, and has to be treated separately. It is absent for $SU(N)$ gauge theories. 
In (\ref{weq1}) $W'$ is the function obtained by differentiating $W(x)$ for some complex number $x$. The following theorem will be proven in the next section
\begin{theorem} \label{tbp}
Equations \ref{weq1} imply the existence of polynomials $H_{N-k},T_{2k},G_{n-k},f_{n-1},U_k$ of degrees $N-k,2k,n-k,n-1,k$, $1\leq k \leq n$ respectively such that
\bea
(W'(M)_+ - W'(M)_-)^2 & = & W'(M)^2 + f_{n-1}(M)  \label{c1} \\
W'(M)_+ - W'(M)_- & = & G_{n-k}(M) (U_k(M)_+ - U_k(M)_-) \label{c2} \\
(P_N(M)_+ - P_N(M)_-)^2 & = & P_N(M)^2 - 4 \Lambda^{2N} \label{c3} \\
P_N(M)_+ - P_N(M)_- & = & H_{N-k}(M) (U_k(M)_+ - U_k(M)_-) \label{c4} \\
(U_k(M)_+ - U_k(M)_-)^2 & = & T_{2k}(M). \label{c5}
\eea
\end{theorem}

In more physical terms, the theorem implies the following factorization of the Seiberg-Witten curve \cite{seibergwittencurvefactorization},
\bea
P_N^2(x) - 4 \Lambda^{2N} & = & H_{N-k}^2(x) T_{2k}(x) \label{mm1}\\
T_{2k}(x) G_{n-k}^2(x) & = & W'^2(x) + f_{n-1}(x).\label{matrixmodelcurve}   
\eea
These equations state that the genus of the Riemann surface is reduced by the introduction of a classical superpotential. Furthermore, these equations show that the matrix model curve (\ref{matrixmodelcurve}) can degenerate as well. 

\section{Algebraic proof}
In this section we will provide an algebraic proof of the above theorem. The first step
of the proof is to show that in addition to (\ref{weq1}) we also have
\be \label{weq2} 
[M,P_N(M)_+ - P_N(M)_-]=0
\ee
or equivalently
\be [\tilde{M},P_N(\tilde{M})_+ - P_N(\tilde{M})_- ]=0 . \ee
This follows from (\ref{defspec}). Indeed, (\ref{defspec}) implies the matrix equation 
\be
P_N(M) + (-1)^N(z+\Lambda^{2N} z^{-1}) = 0
\ee
which when expressed in terms of the infinite matrix $\tilde{M}$ implies
\be
P_N(\tilde{M}) + (-1)^N(Z+\Lambda^{2N} Z^{-1}) = 0.
\ee
Therefore, $P_N(\tilde{M})_+ = (-1)^{N+1} Z$ and $P_N(\tilde{M})_-= (-1)^{N+1} \Lambda^{2N}
Z^{-1}$, and (\ref{weq2}) follows trivially from $[\tilde{M},Z]=0$.

The main idea of the proof is to show that one can compute a ``greatest common
divisor'' of $P_N(M)_+-P_N(M)_-$ and $W'(M)_+ - W'(M)_-$. This computation follows
a version of the Euclid algorithm that one uses compute the greatest common divisor
of two integers. The crucial step is to show that if $[M,A(M)_+-A(M)_-]=0$, then for 
all polynomials $B$ there is a polynomial $C$ such that $C(M)_+-C(M)_- = B(M)
(A(M)_+-A(M)_-)$ and in addition $[M,C(M)_+ - C(M)_-]=0$. We will show this below,
working exclusively with the infinite matrix representation of $M$  (and
dropping the tilde for convenience). 

\subsection{Some definitions}
\label{app:defs}
Let $gl(\infty)_{N}$ denote the elements in $gl(\infty)$ which are periodic with period $N$. 
$X$ is said to be a matrix operator of degree $k$ if
\be
X \in gl(\infty)_{N}, X_{ij} \neq 0 \quad \Longrightarrow \quad i+k=j .
\ee
Any matrix can be written as a sum of matrices of fixed degree, $X=\sum_r X^{(r)}$,
with $X^{(r)}$ of degree $r$. If there is a smallest $r$ for which $X^{(r)}\neq 0$,
we say that $X$ is of minimal degree $r$, and similarly if there is a largest
$r$ for which $X^{(r)}\neq 0$ we say $X$ is of maximal degree $r$.

We will only study operators with a finite maximal degree, which form a ring. By the inverse
of a matrix we will mean an inverse in this ring. If this exists, it is defined
as follows: write a general matrix $R$ of maximal degree $r$ as $R=R^{(r)} + \tilde{R}$,
where $\tilde{R}=\sum_{j<r} R^{(j)}$. $R$ is invertible in the ring if and only
if $R^{(r)}$ is invertible, in which case it is defined as
\be
R^{-1} = (R^{(r)})^{-1} ( 1+ \tilde{R}(R^{(r)})^{-1})^{-1} \equiv
  (R^{(r)})^{-1} \sum_{k\geq 0} (-1)^k ( \tilde{R}(R^{(r)})^{-1})^k
\ee
so that $R^{-1}$ is of maximal degree $-r$.

\subsection{Some facts}
First we will collect a number of intermediate results which will be used to prove the theorem, beginning with the following 
\begin{lemma} \label{lemma1}
Let $X$ be a matrix operator of degree $p\geq 0$.
\be
[X,M^+]=0 \quad {\rm iff} \quad X = \lambda (M^+)^p
\ee 
for some constant $\lambda \in \mathbb{C}$
\end{lemma}

The proof of the first statement follows trivially from the second. 
Clearly, $X=\lambda (M^+)^p$ constitutes a one parameter family of solutions of $[X,M^+]=0$. We will now show that this equation only has a one parameter family of solutions thereby proving the lemma. Write out the equation in components,
\be
[X,M^+]_{ij} = (X_{i,i+p} M^+_{i+p,i+p+1} - M^+_{i,i+1} X_{i+1,i+p+1})\delta_{i+1+p, j}=0 .
\ee
Recall that none of the off-diagonal elements of $M$ can be zero, since $\prod y_i^2 = 
\Lambda^{2N}\neq 0$. Therefore the above equation recursively determines all
entries of $X$ in terms of one arbitrary given entry, and therefore there is a one
complex parameter family of solutions. 

The above lemma is used to construct a proof for
\begin{lemma}\label{lemma2}
Let $X \in gl(\infty)_{N}$ have maximal degree $k$. Then $[X,M]=0$ implies
\be
X = \sum_{-\infty \le i \leq k} \lambda_i M^i .
\ee
\end{lemma}
As before, we denote by
$X^{(p)}$ the component of $X$ of degree $p$. $X$ can be trivially expanded as
\be
X = \sum_{i \leq k} X^{(i)}
\ee
Since $M$ is of maximal degree one, $[X,M]$ is of maximal degree $k+1$, and its
component of highest degree equals
\be
[X,M]^{(k+1)} = [X^{(k)},M^{(1)}]=0.
\ee
Using the previous lemma we have $X^{(k)} = \lambda_k M_+^k$. 
Now consider the operator $X' = X - \lambda_k  M^k$ of maximal degree $k-1$. 
This operator also commutes with $M$. By applying the same procedure
we find $\lambda_{k-1}$ such that $X''=X'-\lambda_{k-1} M^{k-1}$ is of maximal
degree $k-2$. By continuing the procedure the lemma follows. 

By itself the lemma is not that useful yet: we would like to obtain a finite (polynomial) series. Note that the natural operator with a finite series is the upper triangular part of a matrix of finite maximal degree since this matrix can be obtained by the above procedure in a finite number of steps,
\be
X_+ = \left(\sum_{i=1}^k \lambda_i M^i\right)_+
\ee

This observation leads to the following easily proven

\begin{lemma}\label{lemma3}
Let $X \in gl(\infty)_{N}$ have maximal degree $k$ and $[X,M]=0$.
\begin{itemize}
\item $X$ is symmetric then $X = A(M)$
\item $X$ is antisymmetric then $X = A(M)_+ - A(M)_-$
\end{itemize}
with $A(M)$ some polynomial function in M of order $k$
\end{lemma}

Finally we prove the lemma

\begin{lemma}\label{lemma4}
Let $A$ be some some polynomial function in M of order $k$ and $[M, A(M)_+ - A(M)_-]=0$. 
Then $A(M)_0 = \lambda {\mathbb{I}}$ and 
\be
(A(M)_+ - A(M)_-)^2 = A(M)^2 + f_{k-1}(M) + 2 \lambda A(M)
\ee
with $f_{k-1}$ a polynomial function of order $k-1$, $\lambda \in \mathbb{C}$ .
\end{lemma}

Adding $A(M)$ to $A(M)_+ - A(M)_-$ yields $[M_+, A(M)_0]=0$. By lemma \ref{lemma1} $A(M)_0 = \lambda {\mathbb{I}}$. Now consider $(A(M)_+ - A(M)_-)^2$. Since this is a symmetric operator, it is by lemma \ref{lemma3} equal to a polynomial, say $U(M)$. Then

\be
U(M) - A(M)^2 = -2 \lambda A(M) - \lambda^2 - 4 (A_+(M) A_-(M))
\ee

This proves the lemma, since the last term is symmetric and of maximal degree $k-1$.  

\subsection{Proving the theorem}

With these auxiliary results the stage is set for proving the assertions in theorem \ref{tbp}. The fifth assertion is a consequence of lemma \ref{lemma3}. 
The first is a simple application of lemma \ref{lemma4} using the second condition in (\ref{weq1}). This leaves (\ref{c2}), (\ref{c4}). We will find the greatest common divisor indicated in these equation through the following algorithm, which is a variant of the algorithm of
Euclid:

\begin{enumerate}
\item There is a polynomial function $A(M)$ in $M$ of maximum degree $N-n$ such that
\be
(P_N(M)_+ - P_N(M)_-)^{(\pm N)} = A(M)^{(\pm(N-n))}(W'(M)_+ - W'(M)_-)^{(\pm n)}
\ee
\item Define
\be
\tilde{P}(M)_+ - \tilde{P}(M)_- \equiv (P_N(M)_+ - P_N(M)_-) - A(M)(W'(M)_+ - W'(M)_-)
\ee
Note that this operator commutes with $M$ and is of maximum degree $\leq N-1$.
The existence of the polynomial $\tilde{P}$ is guaranteed by lemma 3. 
\item Repeat the above steps with the operators $(W'(M)_+ - W'(M)_-)$ and 
$\tilde{P}(M)_+ - \tilde{P}(M)_-$.
\item At some point the subtraction step will give an
operator of degree zero which commutes with $M$ and is thus proportional to the identity matrix. The other operator is now by construction the greatest common divisor.
\end{enumerate}

Some remarks are in order. First of all, the algorithm stops after a finite number of steps. Furthermore, the maximum degree of the gcd is of course $n$ which is obtained if the algorithm terminates after one step. Denoting the maximum degree of the gcd by $k$ we see we have proven (\ref{c2}) and (\ref{c4}) thereby finishing the proof of theorem \ref{tbp}. 

We have now shown explicitly that the superpotential (\ref{eq:conjecture}) 
indeed reproduces correctly the factorization of the Seiberg-Witten curve for softly broken ${\cal N} = 2 \rightarrow {\cal N} =1$ super Yang-Mills theory \cite{seibergwittencurvefactorization}. This is strong evidence that they give a correct description of the F-terms
of the low-energy effective field theory.  

\section{Algebraic geometrical proof}
\label{sec:alggeomproof}

There is a long literature relating solutions of the equations of motion of the 
periodic Toda chain to algebraic geometric quantities, see e.g. \cite{kacmoerbeke} \cite{datetanaka} \cite{mumfordmoerbeke}.
The basic result that allows
one to do so is that the commuting flows of the Toda system correspond to linear
flows on the Jacobian of the associated spectral curve (\ref{defspec}).
The action variables correspond to moduli of the spectral curve, and are left
invariant by the commuting flows. The angle variables are literally coordinates
on the Jacobian of each of the surfaces. Therefore, the main problem is to
understand the relation between the coordinates on the Jacobian and the variables
that appear in the Lax matrix. This relation involves a set of $g$ points (more precisely,
a divisor of degree $g$) on the spectral curve (of genus $g$).
On the one hand, given $g$ points $P_i$, one other point $P_0$, and a basis of the 
holomorphic one-forms $\omega_i$, the map
\be
\{P_i\} \rightarrow \sum_i \int_{P_0}^{P_i} \omega_j
\ee
maps the $g$ points to a point on the Jacobian, and every point on the Jacobian
can be written in this way via the Jacobi inversion theorem. 
Therefore, in order to understand the relation between the Lax matrix and the
Jacobian, we need to extract $g$ points on the spectral curve from the Lax matrix.
One way to do this is to consider eigenvectors of the Lax matrix
(\ref{deflax}). Viewed as functions on the spectral curve, the entries of the 
eigenvectors generically have poles at $g$ points, which we take to be the $P_i$.
Equivalently, we can consider the spectrum of the matrix $M'$ obtained from
the Lax matrix by removing the last row and column. This is a matrix of
rank $N-1$, and the spectral curve of a generic Lax matrix is indeed a curve
of genus $N-1$. 

It would be interesting to study in more detail the construction and physical
meaning of the points $P_i$, and their behavior under the flows of the Toda lattice,
but we will not do that here. To prove that the Seiberg-Witten curve factorizes in 
the appropriate way, it suffices to use theorem 4 in \cite{mumfordmoerbeke}, which
gives an explicit expression for the velocities of the Toda flows on the Jacobian.
Adapted to our situation, this theorem states that the velocities are given by
\be
a_j = {\rm Res}_{x=\infty} (\omega_j A(x) )
\ee
for the flows given by $\dot{M} = [M,A(M)_+ - A(M)_-]$. Here, $\omega_j$ is
a basis for the holomorphic one-forms on the Riemann surface. For a surface
of the form $y^2=P_N(x)^2 - 4 \Lambda^{2N}$, such a basis is
\be
\omega_j = \frac{x^{j-1}}{\sqrt{P_N(x)^2 - 4\Lambda^{2N}}}, \qquad j=1,\ldots,N-1.
\ee
In order to make the flow given by $A(x)=W'(x)$ stationary, which is according
to (\ref{weq1}) a necessary condition to find an extremum of the superpotential,
we therefore need that
\be \label{aa1}
{\rm Res}_{x=\infty} (\frac{x^{j-1}}{\sqrt{P_N(x)^2 - 4\Lambda^{2N}}} W'(x))=0,
\qquad j=1,\ldots,N-1 .
\ee

Though necessary, these equations are not yet sufficient, since we have not 
yet taken the additional constraint $W'(M)_0$ in (\ref{weq1}) into account. 
This constraint is in fact equivalent to ${\rm Tr}(W'(M))=0$, as one can see
from the proof of lemma 4. Now if $W'$ has a large order, $W'(M)$ appears
to depend explicitly on the spectral parameter, but in our previous paper
we explained what the natural interpretation of such high order $W'$ is, 
namely we should work with the infinite Lax matrix (\ref{inflax}) instead,
and view the trace as the trace over any $N$ consecutive diagonal entries
of the infinite matrix (since the matrix is periodic, this is well-defined). 
With this interpretation of $W'(M)$ for any $W'$, one then obtains that 
${\rm Tr}(W'(M))$ is proportional to
\be \label{aa2}
{\rm Res}_{x=\infty} (\frac{P_N'(x)}{\sqrt{P_N(x)^2 - 4\Lambda^{2N}}} W'(x)).
\ee
Thus, the complete set of equations we need to solve are (\ref{aa1}) 
and (\ref{aa2} = 0), which can be summarized as
\be \label{bb1}
{\rm Res}_{x=\infty} (\frac{x^{j-1}}{\sqrt{P_N(x)^2 - 4\Lambda^{2N}}} W'(x))=0,
\qquad j=1,\ldots,N .
\ee

These equations have no solution if the order of $W'$ is less than $N$.
The only way in which we can have a stationary flow is if the Riemann
surface degenerates, so that the genus becomes smaller and there are fewer 
holomorphic one-forms. So we will assume that the surface degenerates with
\be \label{aux1}
P_N^2(x)-4\Lambda^{2N} = H^2_{N-k}(x) T_{2k}(x)
\ee
in which case the constraints we need to solve read
\be \label{cc}
{\rm Res}_{x=\infty} (\frac{x^{j-1}}{\sqrt{T_{2k}(x)}} W'(x))=0,
\qquad j=1,\ldots,k .
\ee
To obtain this, observe that $P_N'/\sqrt{P_N^2-4\Lambda^{2N}} \rightarrow
V_{k-1}/\sqrt{T_{2k}}$ for some polynomial $V_{k-1}$ if the curve degenerates
according to (\ref{aux1}).
Equations (\ref{cc}) are equivalent to
\be \label{j1}
\frac{W'(x)}{\sqrt{T_{2k}(x)}} = 
G_{n-k}(x)+ \sum_{l > k} \frac{c_l}{x^l}
\ee
with $G_{n-k}$ a polynomial of order $n-k$ (recall that $n$ is the order of $W'$).
Multiplying the left and right hand side by $\sqrt{T_{2k}}$, and taking squares,
shows that 
\be \label{aux2}
W'(x)^2 = T_{2k}(x) G_{n-k}(x)^2 - f_{n-1}(x) 
\ee
where $f_{n-1}(x)$ is a polynomial whose order is at most $n-1$. 
Thus, we have in (\ref{aux1}) and (\ref{aux2}) once again precisely reproduced
the factorization of the Seiberg-Witten curve
(\ref{mm1}) and (\ref{matrixmodelcurve}). 

\section{Comments}

\subsection{Stationary flows}

Although the equations of motion that we obtained from the superpotential imply that
one particular flow is stationary, namely the flow generated by $W(M)$ via
equation (\ref{weq1}), the results in equations (\ref{c1})--(\ref{c5}) show that
$N-k$ inequivalent flows have become stationary. One can take any polynomial $A$
of degree $N-k-1$, and in view of lemma 3 we can always write 
\be
A(M)(U_k(M)_+ - U_k(M)_-) = V(M)_+ - V(M)_-
\ee
for some polynomial $V$, and $M$ commutes with all these operators. Since
$A$ contains $N-k$ free parameters, there are $N-k$ linearly independent
flows that are stationary. 

\subsection{Number of moduli}

We can also easily verify that the Lax matrices that extremize the superpotential
have the right number of moduli that one expects from field theory. At low
energies, for a solution of the form (\ref{c1})--(\ref{c5}), we expect an
unbroken $U(1)^k$ gauge group, and therefore $k$ complex moduli. One of these
is related to the diagonal $U(1)\subset U(N)$ and is not present in the Lax
matrix. Therefore, we expect that the Lax matrix has $k-1$ complex moduli. 
These moduli are easily understood: under the commuting flows, any extremum
of the superpotential is mapped to another extremum, because an extremum is
a stationary point of a particular flow and all flows commute. We already
argued above that $N-k$ flows are stationary, leaving $k-1$ flows that
act non-trivially on the Lax matrix. These are the flows generated by
${\rm Tr}(M^j)$ for $j=2,\ldots,k$. These flows must act non-trivially, 
because if one of them were to act trivially, there would exist a polynomial $V$
of degree less that $k$ such that $[M,V(M)_+-V(M)_-]=0$, and in that case
we would be able to reduce the solution (\ref{c1})-(\ref{c5}) even
further. (We tacitly assumed that $T_{2k}$ has no further quadratic factors in
our solution). The above results are also easily understood in algebraic
geometric framework given above.

\subsection{$SU(N)$ versus $U(N)$}

So far we discussed the theory with gauge group $U(N)$, and it is easy
to see what needs to be changed if we are interested in gauge group $SU(N)$ instead.
The only difference is that we need to drop the additional equation $W'(M)_0=0$ in
(\ref{weq1}), and that we should work with a traceless Lax matrix. In the 
algebraic approach in section~3, we can therefore no longer put $\lambda=0$ when
we apply lemma 4 to $A=W'$, and therefore we end up with precisely the same results
as for $U(N)$, the only difference being that the order of $f$ has to be $n$ 
instead of $n-1$. This is consistent with the fact that ${\rm Tr}(M)=0$, so that
the linear term in $W$ and thus the constant term in $W'$ should have no effect
whatsoever on the factorization and drop out of all equations. A change of the constant
term in $W'$ can be compensated by a change of $f_n$ and indeed does not affect
the physics in any way. 

From the algebraic geometric perspective, changing $U(N)$ to $SU(N)$ implies that
in (\ref{bb1}) and (\ref{cc}) the range of $j$ should be $1\leq j \leq N-1$ and
$1\leq j \leq k-1$ instead. This then implies that we can have a term $c_k/x^k$ 
in equation (\ref{j1}), and therefore $f$ in (\ref{aux1}) can be of order $n$
instead of $n-1$. Thus we reach the same conclusion as above.

\subsection{Outlook}

Given the results in this paper, there are several directions to explore.
What still remains an open problem is the relation between the results in this
paper and the integrable model that underlies the Dijkgraaf-Vafa matrix model
formulation. It would also be interesting to understand whether the various
quantities that appear in the algebraic-geometric formulation have a direct
physical interpretation in the gauge theory. Finally, we hope to extend the
results given here to the other gauge groups, in order to improve our understanding
of the vacuum structure of these theories as well. In \cite{aganagicetc} it was
shown that perturbative calculations reduce to finite dimensional combinatorical
problems for any gauge group and any matter content, but a convenient description
of the vacuum structure of such a general gauge theory is still lacking. The 
three-dimensional analysis may also shed some light on the UV ambiguities discussed
in \cite{aganagicetc}, as they seem closely related to a choice of infinite
realization of a finite dimensional Lax matrix.

\vspace{.5in}

{\bf Acknowledgments: }
We would like to thank Robbert Dijkgraaf for useful discussions. This work is partially supported by the Stichting FOM.

\providecommand{\href}[2]{#2}\begingroup\raggedright\endgroup

%remove blah from following sentence!
%\bibliographyBLAH{3dproof}
%\bibliographystyle{JHEP-2}


\begin{thebibliography}{10}

\bibitem{prevpaper}
R.~Boels, J.~de~Boer, R.~Duivenvoorden and J.~Wijnhout, {\it Nonperturbative
  superpotentials and compactification to three dimensions},
  \href{http://arXiv.org/abs/hep-th/0304061}{{\tt hep-th/0304061}}.
%%CITATION = HEP-TH 0304061;%%

\bibitem{dv1}
R.~Dijkgraaf and C.~Vafa, {\it A perturbative window into non-perturbative
  physics},  \href{http://arXiv.org/abs/hep-th/0208048}{{\tt hep-th/0208048}}.
%%CITATION = HEP-TH 0208048;%%

\bibitem{dv2}
R.~Dijkgraaf and C.~Vafa, {\it N = 1 supersymmetry, deconstruction, and bosonic
  gauge theories},  \href{http://arXiv.org/abs/hep-th/0302011}{{\tt
  hep-th/0302011}}.
%%CITATION = HEP-TH 0302011;%%

\bibitem{seibergwitten3d}
N.~Seiberg and E.~Witten, {\it Gauge dynamics and compactification to three
  dimensions},  \href{http://arXiv.org/abs/hep-th/9607163}{{\tt
  hep-th/9607163}}.

\bibitem{gorsky}
A.~Gorsky, I.~Krichever, A.~Marshakov, A.~Mironov and A.~Morozov, {\it
  Integrability and seiberg-witten exact solution},  {\em Phys. Lett.} {\bf
  B355} (1995) 466--474 [\href{http://arXiv.org/abs/hep-th/9505035}{{\tt
  hep-th/9505035}}].
%%CITATION = HEP-TH 9505035;%%

\bibitem{martinecwarner}
E.~J. Martinec and N.~P. Warner, {\it Integrable systems and supersymmetric
  gauge theory},  {\em Nucl. Phys.} {\bf B459} (1996) 97--112
  [\href{http://arXiv.org/abs/hep-th/9509161}{{\tt hep-th/9509161}}].
%%CITATION = HEP-TH 9509161;%%

\bibitem{nakatsu}
T.~Nakatsu and K.~Takasaki, {\it Whitham-toda hierarchy and n = 2
  supersymmetric yang-mills theory},  {\em Mod. Phys. Lett.} {\bf A11} (1996)
  157--168 [\href{http://arXiv.org/abs/hep-th/9509162}{{\tt hep-th/9509162}}].
%%CITATION = HEP-TH 9509162;%%

\bibitem{dorey1}
N.~Dorey, {\it An elliptic superpotential for softly broken n = 4
  supersymmetric yang-mills theory},  {\em JHEP} {\bf 07} (1999) 021
  [\href{http://arXiv.org/abs/hep-th/9906011}{{\tt hep-th/9906011}}].
%%CITATION = HEP-TH 9906011;%%

\bibitem{dorey2}
N.~Dorey, T.~J. Hollowood and S.~Prem~Kumar, {\it An exact elliptic
  superpotential for n = 1* deformations of finite n = 2 gauge theories},  {\em
  Nucl. Phys.} {\bf B624} (2002) 95--145
  [\href{http://arXiv.org/abs/hep-th/0108221}{{\tt hep-th/0108221}}].
%%CITATION = HEP-TH 0108221;%%

\bibitem{dorey3}
N.~Dorey, T.~J. Hollowood, S.~Prem~Kumar and A.~Sinkovics, {\it Exact
  superpotentials from matrix models},  {\em JHEP} {\bf 11} (2002) 039
  [\href{http://arXiv.org/abs/hep-th/0209089}{{\tt hep-th/0209089}}].
%%CITATION = HEP-TH 0209089;%%

\bibitem{seibergwittencurvefactorization}
J.~de~Boer and Y.~Oz, {\it Monopole condensation and confining phase of n = 1
  gauge theories via m-theory fivebrane},  {\em Nucl. Phys.} {\bf B511} (1998)
  155--196 [\href{http://arXiv.org/abs/hep-th/9708044}{{\tt hep-th/9708044}}].
%%CITATION = HEP-TH 9708044;%%

\bibitem{hollowood}
T.~J. Hollowood, {\it Critical points of glueball superpotentials and
  equilibria of integrable systems},
  \href{http://arXiv.org/abs/hep-th/0305023}{{\tt hep-th/0305023}}.
%%CITATION = HEP-TH 0305023;%%

\bibitem{lax}
P.~D. Lax, {\it Integrals of nonlinear equations of evolution and solitary
  waves},  {\em Comm. Pure Appl. Math.} {\bf 21} (1968) 467--490.

\bibitem{kacmoerbeke}
M.~Kac and P.~van Moerbeke, {\it A complete solution of the periodic {T}oda
  problem},  {\em Proc. Nat. Acad. Sci. U. S. A.} {\bf 72} (1975), no.~8
  2879--2880.

\bibitem{datetanaka}
E.~Date and S.~Tanaka, {\it Analogue of inverse scattering theory for the
  discrete {H}ill's equation and exact solutions for the periodic {T}oda
  lattice},  {\em Progr. Theoret. Phys.} {\bf 55} (1976), no.~2 457--465.

\bibitem{mumfordmoerbeke}
P.~van Moerbeke and D.~Mumford, {\it The spectrum of difference operators and
  algebraic curves},  {\em Acta Math.} {\bf 143} (1979), no.~1-2 93--154.

\bibitem{aganagicetc}
M.~Aganagic, K.~Intriligator, C.~Vafa and N.~P. Warner, {\it The glueball
  superpotential},  \href{http://arXiv.org/abs/hep-th/0304271}{{\tt
  hep-th/0304271}}.
%%CITATION = HEP-TH 0304271;%%

\end{thebibliography}
\end{document}